\documentclass[
superscriptaddress,
bibnotes,
 amsmath,amssymb,
 aps,
pra,
twocolumn,
]{revtex4-1}

\usepackage{graphicx}
\usepackage{dcolumn}
\usepackage{xr}
\usepackage{bm}
\usepackage{hyperref}
\usepackage{color,soul}

\usepackage{lineno}

\begin{document}

\title{Laser cooling of traveling wave phonons in an optical fiber}

\author{Joel N. Johnson}
\address{Department of Applied Physics and Materials Science, Northern Arizona University, Flagstaff, AZ 86011, USA}
\address{Center for Materials Interfaces in Research and Applications, Flagstaff, AZ, USA}
\author{Danielle R. Haverkamp}
\address{Department of Applied Physics and Materials Science, Northern Arizona University, Flagstaff, AZ 86011, USA}
\address{Center for Materials Interfaces in Research and Applications, Flagstaff, AZ, USA}
\author{Yi-Hsin Ou}
\address{College of Optical Sciences, University of Arizona, Tucson, AZ, USA}
\author{Khanh Kieu}
\address{College of Optical Sciences, University of Arizona, Tucson, AZ, USA}
\author{Nils T. Otterstrom}
\address{Photonic and Phononic Microsystems, Sandia National Laboratories, Albuquerque, New Mexico, USA}
\author{Peter T. Rakich}
\address{Department of Applied Physics, Yale University, New Haven, CT, USA}
\author{Ryan O. Behunin}
\email[]{ryan.behunin@nau.edu}
\address{Department of Applied Physics and Materials Science, Northern Arizona University, Flagstaff, AZ 86011, USA}
\address{Center for Materials Interfaces in Research and Applications, Flagstaff, AZ, USA}

\begin{abstract}
In recent years, optical control of mechanical oscillators has emerged as a critical tool for everything from information processing to laser cooling. While traditional forms of optomechanical cooling utilize systems comprised of discrete optical and mechanical modes,  it has recently been shown that cooling can be achieved in a chip-based system that possesses a continuum of modes. Through Brillouin-mediated phonon-photon interactions, cooling of a band of traveling acoustic waves can occur when anti-Stokes scattered photons exit the system more rapidly than the relaxation rate of the mechanical waves---to a degree determined by the acousto-optic coupling. Here, we demonstrate that a continuum of traveling wave phonons can be cooled within an optical fiber, extending this physics to macroscopic length scales. Leveraging the large acousto-optic coupling permitted within a liquid-core optical fiber, heterodyne spectroscopy reveals power-dependent changes in spontaneous Brillouin scattering spectra that indicate a reduction of the thermal phonon population by 21K using 120 mW of injected laser power. 
\end{abstract}

\maketitle


\section{Introduction}
Laser cooling brought about a revolution in atomic, molecular and optical (AMO) physics, permitting exquisite control of the motion of ions, atoms and molecules, precision measurements of time, and the creation of new states of matter \cite{hansch1975cooling,ashkin1978trapping,wineland1978radiation,phillips1982laser,anderson1995observation,davis1995bose,ludlow2015optical,epstein1995observation,sheik2007optical}. 
Beyond these AMO applications, laser cooling has led to impressive developments in solid-state systems where optical refrigeration in rare earth doped glasses is rapidly approaching cryogenic operation, and sideband cooling of individual mesoscopic mechanical oscillators has opened new windows to the foundations of physics and enabled the generation of novel quantum states \cite{chan2011laser,marshall2003towards,hong2017hanbury,aspelmeyer2014cavity}.


Laser cooling of mechanical oscillators conventionally utilize an optical cavity with a movable mirror where radiation pressure mediates a parametric coupling between cavity resonance and a discrete phononic mode \cite{aspelmeyer2014cavity}. This configuration is critical to optical sideband cooling, where red-detuned drive laser photons can blue-shift by phonon scattering, rapidly exit the system, and lower the effective phonon occupancy \cite{aspelmeyer2014cavity}. While optomechanical cooling of traveling wave phonons has been achieved in whispering gallery mode resonators that support discrete optical and mechanical modes \cite{bahl2012observation}, surprisingly, this form of anti-Stokes cooling can occur in continuous systems---without a optical or mechanical resonances \cite{otterstrom2018optomechanical}. Otterstrom {\it et al.} showed that cooling of a continuous band of phonon modes can be achieved when injected laser light scatters through an anti-Stokes process from traveling wave phonons and exits the system more rapidly than the phonon bath returns to thermal equilibrium. This form of cooling simultaneously requires substantial light-sound couplings, to enable efficient scattering, and phonon lifetimes exceeding the transit time for light through the waveguide. Owing to these stringent requirements, optomechanical cooling in continuous systems has only been observed in silicon waveguides engineered to have large Brillouin coupling and short $\sim$ cm lengths. While the realization of this physics within optical fibers may be attractive for low-noise signal processing, high-coherence lasers, high-fidelity optical squeezing, and variable bandwidth forms of all-optical slow-light generation, demonstration of cooling of travelling wave phonons in fibers has remained elusive \cite{shin2015control,shelby1986generation,okawachi2005tunable}.

\begin{figure*}[t]
    \centering
    \includegraphics[width=\textwidth]{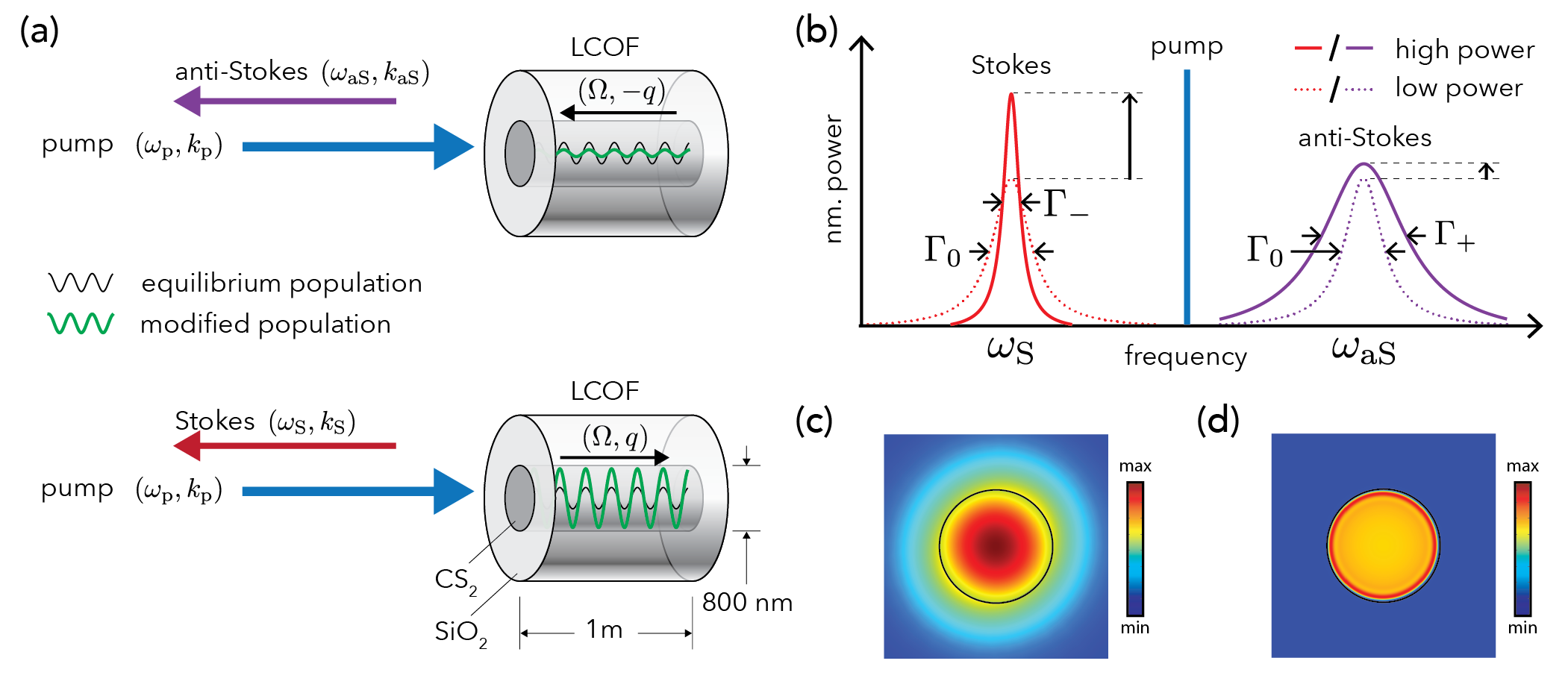}
    \caption{Illustration of liquid core optical fiber (a) geometry and scattering processes, (b) cooling signatures in spontaneous light scattering spectra, and simulated (c) optical and (d) mechanical modes.}
    \label{fig:system}
\end{figure*}

Here, we demonstrate cooling of a continuum of phonon modes in optical fiber for the first time. Through spontaneous Brillouin scattering in a 1m long CS$_2$-filled liquid-core optical fiber, we achieve cooling (i.e., change in temperature) of $\sim$21K from room temperature for a band of anti-Stokes phonons. These dynamics are made possible by the unique features of this fiber system (Fig. \ref{fig:system})\cite{kieu2013brillouin,kieu2014nonlinear,behunin2019spontaneous}. The large refractive index and small sound speed of CS$_2$ enable simultaneous guidance and tight co-confinement of light and sound (Fig. \ref{fig:system}c \& d) within small diameter silica capillaries. This tight acousto-optic overlap, combined with the large electrostrictive coupling made possible within CS$_2$ \cite{boyd2020nonlinear} permit large light-sound interactions which can be leveraged for optomechanical cooling in a continuous system.

Using heterodyne spectroscopy, we show how spontaneously scattered power spectra evolve with pump power. These spectra reveal the signatures of cooling, showing changes in the phonon dissipation rate and nonlinear dependence on the pump power. To confirm depletion of thermal phonons, we perform a form of pump-probe spontaneous Brillouin scattering where changes in the phonon population can be directly probed. This pump-probe technique leverages a unique property of Brillouin scattering, permitting two orthogonal polarizations of light to couple to the same band of phonons. Using this feature an intense pump can be used to cool the phonon modes while a weak probe laser in an orthogonal polarization, that can be isolated from the pump, measures the phonon population. With fixed probe powers, these results show that the power of anti-Stokes scattered probe light decreases as the pump power is increased, indicating that the thermal population of the anti-Stokes phonons has been reduced in a manner consistent with theory. With the long lengths and power-handling capabilities of optical fibers, combined with recent theoretical developments, these results may enable new low-noise fiber applications, ground state cooling of bands of traveling phonon modes, and new forms of quantum state synthesis \cite{zhu2022dynamic,behunin2022quantum}.

\begin{figure*}[t]
    \centering
    \includegraphics[width=\textwidth]{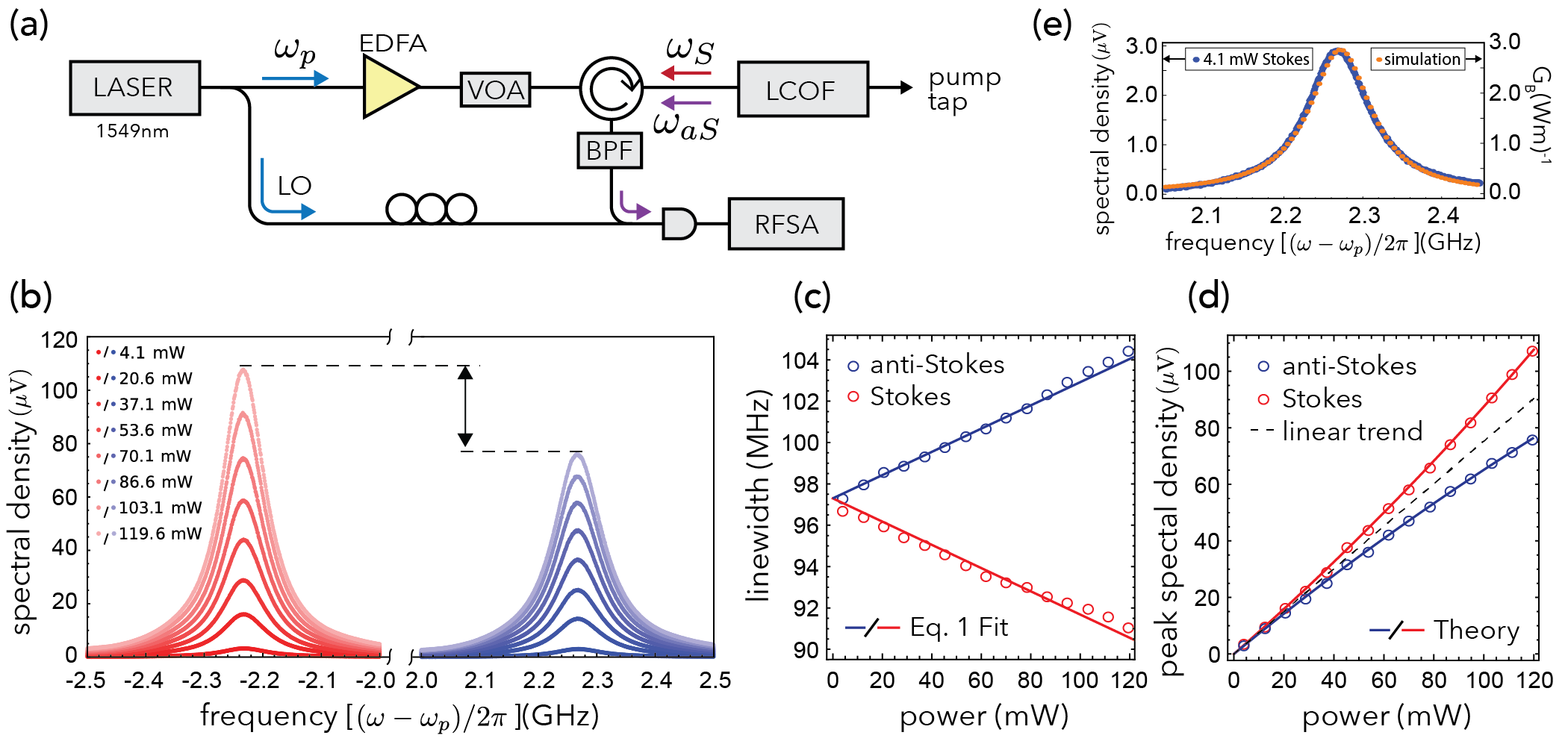}
    \caption{(a) Heterodyne spectroscopy apparatus for measuring power dependence of spontaneous Brillouin spectra shown in (b). Lorentzian fits to spectra in (b) reveal signatures of Brillouin cooling through (c) power-dependent linewidths and (d) sub- and super-linear growth of the power spectra peaks. The lines in (c) are obtained through a constrained fit using Eq. \ref{eq:gamma} providing an estimated Brillouin gain of 2.3 (Wm)$^{-1}$. Theory in (d) obtained using the fitted value of $G_B$, Eq. \eqref{eq:neff} and the measured spectra heights at 4.1 mW (see Appendix A).(e) Comparison between measured and simulated light scattering spectra (see Appendix C). Here, EDFA stands for Erbium doped fiber amplifier, BPF is a tunable band-pass filter, and VOA is variable optical attenuator.}
    \label{fig:results-1}
\end{figure*}

\section{Laser cooling of traveling wave phonons} 
Laser cooling of traveling wave phonons occurs through anti-Stokes Brillouin scattering. During this process an incident pump photon 
annihilates a counter-propagating phonon, blue-shifts to higher energy, and rapidly exits the system (Fig. \ref{fig:system}a). In order for cooling to occur, the mechanical degrees of freedom must return to thermal equilibrium more slowly than the anti-Stokes photons exit the system. A mean-field analysis of these dynamics, described in Appendix A, shows that these conditions require $4v_g/L > \Gamma_0$, where $v_g$ is the optical group velocity, $L$ is the system length and $\Gamma_0$ is the mechanical dissipation rate \cite{otterstrom2018optomechanical}, conditions met by the system considered here (Fig. \ref{fig:system}(a)) with $4 v_g/L \approx 0.82$ GHz and $\Gamma_0 \approx 0.61$ GHz. 
While these disparate optical and mechanical timescales allow the phonon population to be driven out of thermal equilibrium, the degree of cooling requires a relatively large single pass Brillouin gain $G \equiv G_B P_p L$ where $G_B$ is the Brillouin gain coefficient [m$^{-1}$W$^{-1}$] and $P_p$ is the pump power \cite{boyd2020nonlinear}, yielding a fractional temperature $\Delta T/T_0$ change given approximately by 
\begin{equation}
\label{eq:FFC}
   \frac{\Delta T}{T_0} \approx \frac{G}{G+4} 
\end{equation} 
where $T_0$ is the equilibrium temperature \cite{otterstrom2018optomechanical}. Therefore, we must balance competing requirements, maximizing the single-pass gain while minimizing the light travel time across the fiber to achieve efficient cooling. Although the mean-field optical decay rate is marginally greater than the phonon decay rate, analysis of the coupled envelope dynamics shows that the key conclusions of the mean-field analysis remain valid in the limit of small single-pass gains explored here (see Appendix A.1).  

Cooling of traveling wave phonons can be identified by power-dependent changes in the width and height of spontaneous light scattering spectra, revealing the phonon band temperature and dissipation rate (Fig. \ref{fig:system}b). While for low power ($G \ll 1$) the spectrum height increases linearly with pump power, these increases begin to saturate (grow more rapidly) for the anti-Stokes (Stokes) spectrum as the single pass gain approaches one and the phonon population is depleted (increased). Consequently, the anti-Stokes (Stokes) spectrum height grows sub-linearly (super-linearly), depending on both the pump power and the phonon population (Figs. \ref{fig:system}b \& \ref{fig:results-1}d). Owing to additional decay (amplification) channels opened by spontaneous anti-Stokes (Stokes) Brillouin scattering, the phonon dissipation rate increases (decreases) with pump power according to 
\begin{equation}
\label{eq:gamma}
   \Gamma_{\pm} \approx \Gamma_0\left(1\pm \frac{1}{4}G\right)
\end{equation}
where the upper (+) sign indicates an increase in the anti-Stokes phonon decay rate and the lower (-) sign quantifies how the Stokes spectrum narrows \cite{otterstrom2018optomechanical}. As a consequence of these altered decay rates, detailed balance for thermal equilibrium is broken, reducing (or increasing) the effective population of the anti-Stokes (Stokes) phonons $n^+_{eff}$($n^-_{eff}$) described by 
\begin{equation}
\label{eq:neff}
    n^{\pm}_{eff} \approx \frac{\Gamma_0}{\Gamma_{\pm}} n_{th},
\end{equation}
where $n_{th} \approx 2700$, given by the Bose distribution, is the occupation number for the relevant phonons in our liquid-core optical fiber (LCOF) in thermal equilibrium \cite{otterstrom2018optomechanical}. See Appendix A
for more details of the mean-field analysis of the nonlinear optics in this system. 

To demonstrate this physics, we utilize a 1m long CS$_2$-filled silica capillary (Fig. \ref{fig:system}), possessing an array of properties that are ideal for laser cooling of travelling wave phonons. The materials properties and geometry of this liquid-filled capillary permit guided optical (single-mode at 1.55 $\mu$m) (Fig. \ref{fig:system}c) and acoustic waves (Fig. \ref{fig:system}d) within the fiber core and enable large Brillouin gain. Furthermore, macroscopic meter-scale lengths make large single-pass gain accessible at modest sub-Watt powers. See Refs. \cite{kieu2013brillouin,kieu2014nonlinear,behunin2019spontaneous} for details on Brillouin scattering in LCOFs. 

\section{Methods \& Results}
We measure Brillouin cooling in this LCOF using the apparatus shown in Fig. \ref{fig:results-1}a. The output of a laser is split in two, a variable power arm is injected into the sample for Brillouin cooling, and a fixed-power arm is used to synthesize a local oscillator (LO) for heterodyne detection. Spontaneously backscattered light is filtered, to remove stray pump light and isolate the anti-Stokes (or Stokes) sideband, photomixed with the LO on a high-speed receiver, and detected on a radio-frequency spectrum analyzer (RFSA) as a function of pump power. These measurements yield spontaneous Brillouin scattering spectra of the type shown in Fig. \ref{fig:results-1}b.

\begin{figure*}[t]
    \centering \includegraphics[width=0.95\textwidth]{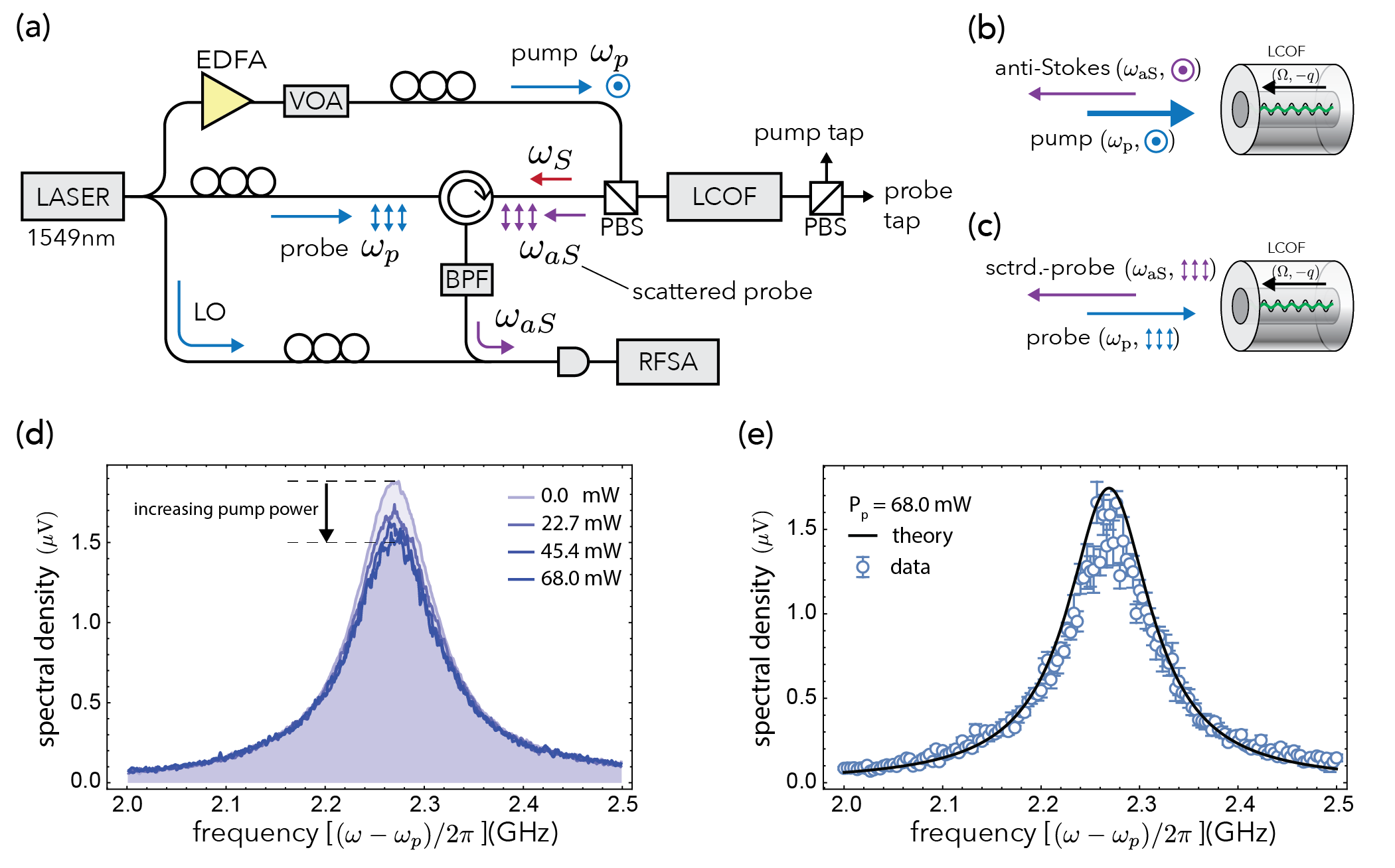}
    \caption{Pump-probe measurements of Brillouin cooling. (a) Pump-probe spectrometer. Orthogonally polarized pump and fixed-power probe light are combined and sent into the sample using a polarizing beamsplitter (PBS). Backscattered pump and probe light exit distinct ports of the PBS, and anti-Stokes sideband is isolated using a circulator and tunable bandpass filter (BPF). Orthonally polarized pump and probe beams (b) and (c) couple to the same band of phonons. The scattered probe can be isolated from the scattered pump using a PBS. 
    (d) Backscattered anti-Stokes power spectra for the probe plotted for various pump powers. For clarity the data is averaged over 1.67 MHz bins. (e) Comparison of measured anti-Stokes spectra and the power spectrum predicted by mean-field theory (see Appendix B).}  
    \label{fig:results-2}
\end{figure*}

The data displayed in Fig. \ref{fig:results-1} exhibit key signatures of Brillouin cooling. With increasing pump power, the linewidth of the anti-Stokes spectrum increases and the scattered anti-Stokes power increases sub-linearly (Fig. \ref{fig:results-1}c \& d). The extracted linewidths of the Stokes and anti-Stokes spectra exhibit the power dependence described by Eq. \eqref{eq:gamma}, yielding a Brillouin gain coefficient $G_B \sim 2.3$(Wm)$^{-1}$, qualitatively described by finite-element simulations shown in Fig. \ref{fig:results-1}e predicting $G_B \approx 2.9$ (Wm)$^{-1}$. 
Using Eq. \eqref{eq:FFC}, these measurements show that the temperature of the band of traveling anti-Stokes phonons has been reduced by 21K from room temperature for 120mW of pump power.

The apparatus and data shown in Fig. \ref{fig:results-1} directly demonstrate the signatures of laser cooling of a continuous band of phonons. However, in these measurements, the amplitude of the power spectrum is proportional to the product of phonon population and the pump power. To directly detect changes to the phonon population, we utilize a form of pump-probe spectroscopy shown in Fig. \ref{fig:results-2}. In contrast with the heterodyne measurements depicted in Fig. \ref{fig:results-1}, the output of a laser is divided into a variable power pump, and a fixed power probe. Orthogonally polarized pump and probe beams are combined using a polarizing beam splitter (PBS) and sent into the LCOF where they interact with the same phonon mode (Fig. \ref{fig:results-2} b \& c). Owing to this polarization multiplexing, spontaneously back-scattered probe light is separated from the backscattered pump by the PBS, filtered, and photomixed with the LO for heterodyne detection, yielding the spectra shown in Fig. \ref{fig:results-2}d. Consistent with the depopulation of phonons, these results show a decrease in the anti-Stokes scattering rate with increasing in pump power. Furthermore, Fig. \ref{fig:results-2}c shows that these spectra are well-described by a simple model of the pump-probe dynamics with parameters obtained from independent measurements shown in Fig. \ref{fig:results-1}. See Appendix B for further details of the pump-probe theory. 

\section{Conclusions}
We have demonstrated laser cooling of travelling wave phonons in an optical fiber for the first time.  Using spontaneous Brillouin light scattering spectroscopy of a CS$_2$-filled LCOF, we probe the populations of travelling wave phonons, revealing power-dependent changes in the phonon dynamics consistent with a mean-field model and finite-element simulations. Key to accessing this regime of light matter interactions is the large acousto-optic coupling and short length of our sample, where pump photons readily scatter from counter-propagating phonons and rapidly exit the fiber before the phonons can return to thermal equilibrium. The combination of these properties enables 21K of phonon cooling for 120mW of pump power. 

Building on these results, drastic enhancements in our ability to cool traveling wave phonons may be achievable with changes in the fiber geometry. With larger core diameters, the Brillouin gain can be dramatically increased through improved acousto-optic overlap. Working toward the limit of single-mode operation (for 1.55 $\mu$m optical wavelengths), finite-element simulations predict Brillouin gains of $\sim$ 11 (Wm)$^{-1}$ for a LCOF with a 1.6 $\mu$m core diameter, exceeding the acousto-optic coupling in our LCOF by nearly a factor of five. Under the same experimental conditions as this study, such a fiber could achieve 119 K of cooling for 250 mW of pump power, approaching cryogenic temperatures. Because thermal noise produced by traveling wave phonons is a critical limitation in a variety of fiber-based information processing, lasers, and quantum optics experiments, these results may inspire new ways to achieve improved performance.  

\section{Acknowledgements}

J. J. and R. B. acknowledge funding from NSF Award No. 2145724. This article has been authored by an employee of National Technology \& Engineering Solutions of Sandia, LLC under Contract No. DE-NA0003525 with the U.S. Department of Energy (DOE). The employee owns all right, title and interest in and to the article and is solely responsible for its contents. The United States Government retains and the publisher, by accepting the article for publication, acknowledges that the United States Government retains a non-exclusive, paid-up, irrevocable, world-wide license to publish or reproduce the published form of this article or allow others to do so, for United States Government purposes. The DOE will provide public access to these results of federally sponsored research in accordance with the DOE Public Access Plan https://www.energy.gov/downloads/doe-public-access-plan.

 


\appendix

\section{Mean-field analysis of Brillouin cooling}
\label{SI:meanField}
The key features of Brillouin cooling can be understood from a mean-field analysis of the slowly-varying envelope equations. For an undepleted pump and ignoring the propagation of phonons, spontaneous Brillouin scattering can be described by the following coupled stochastic envelope equations 
\begin{eqnarray}
\label{SVE}
   && \dot{A}_S + v_g \partial_z A_S = -i g A_p B_S^\dag \\
   && \dot{B}_S + \frac{\Gamma_0}{2} B_S  = -i g A_p A_S^\dag + \xi_S \\
      \label{SVE-3}
   && \dot{A}_{aS} + v_g \partial_z A_{aS} = -i g A_p B_{aS} \\
   \label{SVE-4}
   && \dot{B}_{aS} + \frac{\Gamma_0}{2} B_{aS}  = -i g A_{aS} A_p^\dag + \xi_{aS}
\end{eqnarray}
expressed in a frame rotating at the resonance frequency for each field \cite{kharel2016noise}.
Here, $A_{S}$ ($A_{aS}$) and $B_{S}$ ($B_{aS}$) are the respective (right-moving) optical and mechanical envelopes for the Stokes (anti-Stokes) process, $A_p$ is the envelope for a constant undepleted pump, $g$ is the Brillouin coupling, and $v_g$ is the optical group velocity. Thermal fluctuations of the mechanical field are modeled using the locally correlated white noise Langevin forces 
$\xi_{j}$ ($j=S$ or $aS$)  with correlation properties
\begin{eqnarray}
  &&  \langle \xi_j(t,z) \xi_{j'}^\dag(t',z')\rangle = \Gamma_0 (n_{th}+1) \delta_{jj'} \delta(t-t')\delta(z-z') \quad \quad \\
  \label{corr-2}
  && \langle \xi_j^\dag(t,z) \xi_{j'}(t',z')\rangle = \Gamma_0 n_{th} \delta_{jj'} \delta(t-t')\delta(z-z')
\end{eqnarray}
and $\Gamma_0$ is the mechanical dissipation rate. Put in terms of readily measurable quantities, the power in an optical mode is given by $P_{S/aS/p} = \hbar \omega_{S/aS/p} v_g A^\dag_{S/aS/p}A_{S/aS/p}$ and the Brillouin gain can be expressed as 

\begin{equation}
\label{eq:GB}
    G_B = \frac{4 |g|^2}{\hbar \omega_p v_g^2 \Gamma_0}.
\end{equation}

We define the mean fields for the respective optical and mechanical fields $a_j$ and $b_j$  as the average of the envelope over the length of the waveguide
\begin{eqnarray}
&&  a_j = \frac{1}{L}\int_0^L dz \ A_j(z) 
\\
&&  b_j = \frac{1}{L}\int_0^L dz \ B_j(z).
\end{eqnarray}
 
 The mean field equations can be obtained by averaging Eqs. \eqref{SVE}-\eqref{SVE-4} over the waveguide length, giving 
\begin{eqnarray}
\label{eq:mean-field}
   && \dot{a}_S + \frac{\gamma}{2}a_S = -i g A_p b_S^\dag \\
   \label{eq:mean-field-2}
   && \dot{b}_S + \frac{\Gamma_0}{2} b_S  = -i g A_p a_S^\dag + \bar{\xi}_S \\
   && \dot{a}_{aS} + \frac{\gamma}{2} a_{aS} = -i g A_p b_{aS} \\
   \label{eq:mean-field-4}
   && \dot{b}_{aS} + \frac{\Gamma_0}{2} b_{aS}  = -i g a_{aS} A_p^\dag + \bar{\xi}_{aS}
\end{eqnarray}
where $\bar{\xi}_j$ is the spatial average of the Langevin force. 
Here, the mean-field optical decay rate $\gamma = 4 v_g/L$ is obtained from the spatial average of the derivative terms, assuming $a_j \approx (A_j(L)+A_j(0))/2$ and using that the scattered optical fields vanish at the input face, i.e., $A_j(0) = 0$, according to
\begin{eqnarray}
    \frac{v_g}{L} \int_0^L dz  \ \partial_z A_k(z)  = && \frac{v_g}{L}(A_k(L)-A_k(0)) \nonumber
   \\
    = && \frac{v_g}{L}(A_k(L)+ A_k(0)-2A_k(0))  \nonumber
   \\
    = && \frac{2v_g}{L}a_k.
\end{eqnarray}

Next we analyze the mean field dynamics. When $\gamma > \Gamma_0$, the optical fields respond to changes in the phonon amplitude faster than the phonon changes, enabling a quasistatic solution to the optical field dynamics given by 
\begin{eqnarray}
\label{eq:Ad-Elim}
   &&  \frac{\gamma}{2}a_S \approx -i g A_p b_S^\dag \\
   \label{eq:Ad-Elim-2}
   &&  \frac{\gamma}{2} a_{aS} \approx -i g A_p b_{aS}.
\end{eqnarray}
Inserting Eq. \eqref{eq:Ad-Elim} \& \eqref{eq:Ad-Elim-2} into Eqs. \eqref{eq:mean-field-2} \& \eqref{eq:mean-field-4} we find the effective phonon dynamics given by 
\begin{eqnarray}
\label{eq:eff-mean-field}
   && \dot{b}_S + \frac{1}{2} \Gamma_S b_S  =  \bar{\xi}_S \\
   \label{eq:eff-mean-field-2}
   && \dot{b}_{aS} + \frac{1}{2}\Gamma_{aS} b_{aS}  =  \bar{\xi}_{aS}
\end{eqnarray}
where $P_p = \hbar \omega_p v_g A^\dag_p A_p$, Eq. \eqref{eq:GB} and Eq. (2) of the main text have been used. In addition, to changing the effective phonon dynamics, these results show that the power spectra for the optical fields are proportional to the phonon power spectrum, 
\begin{eqnarray}
   && \!\!\!\!\!\!\!\!\!\!S_S[\omega] = \frac{4|g|^2 |A_p|^2}{\gamma^2} \!\int_{-\infty}^{\infty} d\tau \ e^{i\omega \tau} \langle b_S(t+\tau) b_S^\dag(t)\rangle 
    \\
   && \!\!\!\!\!\!\!\!\!\! S_{aS}[\omega] = \frac{4|g|^2 |A_p|^2}{\gamma^2} \!\int_{-\infty}^{\infty} d\tau \ e^{i\omega \tau} \langle b^\dag_{aS}(t\!+\!\tau) b_{aS}(t)\rangle \quad \quad 
\end{eqnarray}
illustrating how the optical power spectra, defined by
$S_j[\omega] = \int_{-\infty}^{\infty} d\tau \ \exp\{i\omega\tau\} \langle a^\dag_{j}(t+\tau) a_{j}(t)\rangle$, permit a form of nonequilibrium phonon spectroscopy. Solving Eqs. \eqref{eq:eff-mean-field} and using the correlation properties for the spatially averaged Langevin forces given by
\begin{eqnarray}
  &&  \langle \bar{\xi}_j(t) \bar{\xi}_{j'}^\dag(t')\rangle = \frac{1}{L}\Gamma_0 (n_{th}+1) \delta_{jj'} \delta(t-t') \\
  && \langle \bar{\xi}_j^\dag(t) \bar{\xi}_{j'}(t')\rangle = \frac{1}{L}\Gamma_0 n_{th} \delta_{jj'} \delta(t-t'),
\end{eqnarray}
we find
\begin{eqnarray}
   && S_S[\omega] = \frac{\Gamma_0 G}{4  v_g} \frac{\Gamma_0 n_{th}}{\Gamma_S} \frac{\Gamma_S}{\omega^2 + \Gamma_S^2/4} 
    \\
   && S_{aS}[\omega] = \frac{\Gamma_0 G }{4 v_g} \frac{\Gamma_0 n_{th}}{\Gamma_{aS}} \frac{\Gamma_{aS}}{\omega^2 + \Gamma_{aS}^2/4}.
\end{eqnarray}
The peak of the power spectrum (i.e., on resonance, or $\omega = 0$), used for the theory in Fig. 2e, is given by 
\begin{eqnarray}
   && S^{peak}_S[P_p] = \frac{n_{th}}{v_g} \frac{GB P_p L}{(1-\frac{1}{4} G_B P_p L)^2} 
    \\
   && S^{peak}_{aS}[P_p] = \frac{n_{th}}{v_g} \frac{G_B P_p L}{(1+\frac{1}{4} G_B P_p L)^2}.
\end{eqnarray}
To account for the radio-frequency to electrical conversion, the peak of the power spectrum at the lowest power ($P_0 = 4.1$ mW) is used to scale the theoretical curves to have units of $\mu V$
\begin{eqnarray}
    && S^{peak,RF}_S[P_p] = \frac{S^{peak,RF}_S[P_0]}{G_B P_0 L} \frac{G_B P_p L}{(1-\frac{1}{4} G_B P_p L)^2} \quad \quad
    \\
   && S^{peak,RF}_{aS}[P_p] = \frac{S^{peak,RF}_{aS}[P_0]}{G_B P_0 L} \frac{G_B P_p L}{(1+\frac{1}{4} G_B P_p L)^2}. \quad\quad\quad
\end{eqnarray}

\subsection{Envelope analysis of cooling dynamics, and validity of the mean-field model}
Here, we show that in the limit of small single-pass gain, the conclusions of the mean-field model agree with the full envelope dynamics.

In the Fourier domain, Eq. \eqref{SVE-4} can be readily solved and inserted in Eq. \eqref{SVE-3}, giving

\begin{eqnarray}
\label{SVE-3p}
   && (\partial_z -i \Lambda)A_{aS}(\omega,z) = -\frac{1}{v_g}i g A_p \hat{B}_{aS}(\omega,z) \quad
\end{eqnarray}
where 
\begin{eqnarray}
    \Lambda = \frac{1}{v_g}\bigg[\omega + i \frac{|g|^2 |A_p|^2}{
-i\omega + \Gamma_0/2} \bigg],
\end{eqnarray}
and \begin{eqnarray}
\label{Bhat}
\hat{B}(\omega,z) = \frac{1}{
-i\omega + \Gamma_0/2} \xi(\omega,z).
\end{eqnarray}
Using Eqs. \eqref{corr-2} and \eqref{Bhat}, the solution to Eq. \eqref{SVE-3p} at the exit face of the fiber ($z=L$), 
\begin{eqnarray}
    A_{aS}(\omega,L) = -i \frac{g A_p}{v_g}\int_0^{L} dz \ e^{i\Lambda(L-z)} \hat{B}(\omega,z), 
\end{eqnarray}
can be directly used to obtain the power spectrum of spontaneously scattered anti-Stokes light
\begin{eqnarray}
  S_{aS,env}[\omega] &\equiv& \frac{\langle A_{aS}^\dag(\omega,L) A_{aS}(\omega',L)\rangle}{2\pi\delta(\omega-\omega')} \nonumber \\
  &=& \frac{n_{th}}{v_g}
  \bigg[ 
  1-e^{
  -G(\omega)} \bigg]
\end{eqnarray}
where
\begin{eqnarray}
   G(\omega) =  \frac{\Gamma^2/4}{\omega^2+\Gamma^2/4} G.
\end{eqnarray}
Noting that the power spectrum obtained directly from \eqref{SVE-3} is directly related to the effective phonon power spectrum $S_B[\omega]$
\begin{eqnarray}
 S_{aS,env}[\omega] \!\!\!& = &\!\!\! \left| \frac{g A_p}{v_g} \right|^2 \!\! \int_0^{L} \!\!\!\!\! dz\!\!\int_0^{L} \!\!\!\!\!dz'
 e^{i\frac{\omega}{v_g}(z\!-\!z')}
 \frac{\langle {B}^\dag(\omega,z) {B}(\omega',z')\rangle}{2\pi\delta(\omega-\omega')} \nonumber \\
& = &\!\!\! \left| \frac{g A_p}{v_g} \right|^2 S_B[\omega],
\end{eqnarray}
(note the power spectrum of $B$ as opposed to $\hat{B}$) we identify the anti-Stokes phonon power spectrum including the effects of optomechanical cooling
\begin{eqnarray}
\label{phonon_ps}
 S_B[\omega] = \frac{4 n_{th} L}{G \Gamma_0}(1-e^{-G(\omega)}).
\end{eqnarray}
Using $S_B[\omega]$ we can calculate the effective thermal occupation of the anti-Stokes phonon mode
\begin{eqnarray}
 n_{eff,env} & = & \frac{1}{2\pi L}\int_{-\infty}^{\infty} d \omega \ S_B[\omega] 
 \\
 & = & n_{th} e^{-G/2}(I_0(G/2) + I_1(G/2))
\end{eqnarray}
where the $\omega$-integral can be expressed in terms of modified Bessel function $I_0$ and $I_1$. Moreover, the full-width at half maximum $\Gamma_{aS,env}$ can be calculated from $S_B$ using $S_B[\Gamma_{aS,env}/2]= S_B[0]/2$ giving effective anti-Stokes phonon decay rate derived from the envelope dynamics given by  
\begin{eqnarray}
    \Gamma_{aS,env} = \Gamma_0\left[ \frac{G}{-\ln\left((1+e^{-G})/2\right)}-1 \right]^{1/2}.
\end{eqnarray}
In the limit of small single-pass gain $G$, these results can be compared with the mean-field model, showing agreement to order $G$ for the effective phonon decay rate and order $G^2$ for the effective thermal occupation
\begin{align}
& \!\!\!\! n_{eff}  \approx n_{th}(1-G/4+G^2/16- G^3/64 + ...) \\
& \!\!\!\! n_{eff,env}   \approx  n_{th}(1-G/4+G^2/16- 5G^3/384 + ...)  \\ 
& \!\!\!\! \Gamma_{aS}  =  \Gamma_0(1+G/4) \\
& \!\!\!\! \Gamma_{aS,env}  \approx  \Gamma_0(1+G/4+G^2/32+...).
\end{align}
    For the maximum single pass gain explored in this study $G < 0.3$, the results of the mean-field and envelope models agree to better than $0.25 \%$ in effective lifewidths and better than $0.007\%$ in effective thermal occupation. The above analysis is similar for the Stokes process. 

\section{Pump-probe theory}
\label{SI:pump-probe}
In this section, we develop a mean-field theory to describe the pump-probe measurements. In these measurements, both pump and probe couple to the same phonon yielding the coupled envelope equations given by
\begin{eqnarray}
\label{eq:pump-probe}
   && \dot{a}_{aS} + \frac{\gamma}{2} a_{aS} = -i g A_p b_{aS} \\
   && \dot{a}_{sig} + \frac{\gamma}{2} a_{sig} = -i g A_{pr} b_{aS} \\
   && \dot{b}_{aS} + \frac{\Gamma_0}{2} b_{aS}  = -i g a_{aS} A_p^\dag -i g a_{sig} A_{pr}^\dag + \bar{\xi}_{aS} \quad \quad
\end{eqnarray}
where $A_{aS}$ is the anti-Stokes light scattered from the pump $A_p$ and $A_{sig}$ is the anti-Stokes light scattered by the probe the undepleted probe laser $A_{pr}$. 

In the quasistatic limit (i.e., $\gamma > \Gamma_0$) and assuming the $|A_p| \gg |A_{pr}|$, the effective phonon dynamics is given by Eq. \eqref{eq:eff-mean-field-2} and the power spectrum of the scattered probe light is given by  
\begin{eqnarray}
\label{eq:pump-probe-power}
    S_{sig}[\omega,P_p] && \equiv \int_{-\infty}^\infty d\tau \ e^{i\omega\tau}\langle a^\dag_{sig}(t+\tau) a_{sig} (t) \rangle \nonumber  \\
   && = \frac{\Gamma_0 G_p L P_{pr} }{4 v_g} \frac{\Gamma_0 n_{th}}{\Gamma_{aS}} \frac{\Gamma_{aS}}{\omega^2 + \Gamma_{aS}^2/4}.
\end{eqnarray}
where $P_{pr}$ is the fixed probe power and $\Gamma_{aS}$ is the phonon decay rate dependent upon the pump power defined in Eq. (2) of the main text. This result shows that as $\Gamma_{aS}$ increases, i.e., with increased pump power, the power spectrum broadens and decreases in magnitude. To obtain the theoretical curve shown in Fig. 2c, we use the peak value of $S_{sig}[0,0]$ to determine the constant prefactor in Eq. \eqref{eq:pump-probe-power} where $\Gamma_{aS} \to \Gamma_0$, including the optical to radiofrequency conversion
\begin{equation}
    S_{sig}[0,0] = \frac{ G_p L P_{pr} }{ v_g} \frac{n_{th}}{\Gamma_{0}^2}.
\end{equation}
Moving out of the rotating frame yields the prediction for the scattered probe power spectrum is
\begin{eqnarray}
\label{eq:pump-probe-power-RF}
    S^{RF}_{sig}[\omega,P_p] = \frac{\Gamma_{0}^2/4}{(\omega-\omega_{pr}-\Omega)^2 + \Gamma_{aS}^2/4} S^{RF}_{sig}[0,0],
\end{eqnarray}
and used to plot the theory in Fig. 2e.

\section{Brillouin gain simulations}

We calculate the Brillouin gain spectrum by utilizing finite element simulations of the optical and acoustic modes of the LCOF. By including empirically derived material properties (see Tab. \ref{parameters}) and damping for silica \cite{vacher1981ultrasonic} and CS$_2$ \cite{coakley1975brillouin}, these simulations yield the spatial mode profiles for the electric field ${\bf E}$ and the elastic displacement ${\bf u}$ (Fig. 1c \& d). We obtain the Brillouin gain by evaluating the dissipated mechanical power using the equation
\begin{equation}
G_{B} = \frac{\omega_{\rm p}}{\Omega} \frac{1}{P_p P_S} \int_{wg} d^2x \ \langle {\bf f} \cdot \dot{\bf u} \rangle
\end{equation}
where $P_p$ and $P_S$ are the respective pump and Stokes powers of the simulated electromagnetic fields, $\Omega$ is the angular frequency of the mechanical mode, ${\bf f}$ is the electrostrictive force density, $\int_{wg}$ is an integral over the waveguide cross-section, and  $\langle  {\bf A}\cdot {\bf B} \rangle$ is the time average of the vector product ${\bf A}\cdot {\bf B}$. For propagation along the z-axis,  the electrostrictive force density ${\bf f}$ is given by 
\begin{align}
\label{ }
& f_j = \frac{1}{2} \varepsilon_0 n^4 p_{ijkl} \partial_i(E_k E_l^*)
\quad {\rm (silica)}
\\
& {\bf f} = - \frac{1}{4} \varepsilon_0 \gamma_e \nabla |{\bf E}|^2 \quad ({\rm CS}_2)
\end{align}
where $\varepsilon_0$ is the permittivity of vacuum, $n$ is the refractive index, $p_{ijkl}$ is the photoelastic tensor, $\gamma_e$ is the electrostrictive constant, and the Einstein summation convention is assumed for repeated indices.  
In these expressions, we neglect the modal differences between the pump and the Stokes fields.

Acoustic dissipation critically determines the predicted power spectrum. We account for dissipation by including empirically obtained acoustic quality factors for silica and CS$_2$ in our simulations \cite{coakley1975brillouin,vacher1981ultrasonic}. The parameters used in our simulation are summarized in the table below.

\begin{table}[ht]
\begin{tabular}{|c|c|c|c|}
\hline material   & parameter 		& value 		            \\ 
\hline CS$_2$ 	& density      		&  1260   kg/m$^3$  	    \\
 		& refractive index 		&  1.5885		              \\
 		& speed of sound   		&  1226     m/s	             \\
		&electrostrictive constant ($\gamma_e$)  & 2.297   \\ 
  & acoustic quality factor & 23.5 \\
\hline SiO$_2$ 	&density  		     	&  2203     kg/m$^3$	     \\ 
 		& refractive index  		&  1.445    		    \\
 		& Young's modulus		&  73.1     GPa		    \\
 		& shear modulus   		&  31.24    GPa		    \\
 		& phot. elas. tensor (p$_{11}$,p$_{12}$,p$_{44}$) & (0.125,0.27,-0.073)   \\ 
   & acoustic quality factor & 1800 \\
\hline
\end{tabular} 
\centering
\caption{Parameters used in simulations of spontaneous Brillouin scattering spectra.}
\label{parameters}
\end{table}

\bibliography{ref}

\end{document}